\documentclass[twocolumn,prl,aps,showpacs,longbibliography,10pt]{revtex4-1}
\usepackage{graphicx}
\usepackage{natbib}
\usepackage{mathcomp}
\usepackage{url}
\urldef\cmcdata\url{http://kcdb.bipm.org/AppendixC/country_list_search.asp?CountSelected=GB&iservice=EM/DC.2.1.3}
\setcitestyle{nature}

\begin{document}

\title{Towards a quantum representation of the ampere using single electron pumps} 

\author{S.~P.~Giblin,$^{1}$ M. Kataoka,$^{1}$ J.~D.~Fletcher,$^{1}$ P.~See,$^{1}$ T.~J.~B.~M.~Janssen,$^{1}$
J.~P.~Griffiths,$^{2}$ G.~A.~C.~Jones,$^{2}$ I.~Farrer,$^{2}$ and D.~A.~Ritchie$^{2}$}

\affiliation{$^{1}$ National Physical Laboratory, Hampton Road, Teddington, Middlesex TW11 0LW, United Kingdom }

\affiliation{$^{2}$ Cavendish Laboratory, University of Cambridge, J J Thomson Avenue, Cambridge CB3 0HE, United Kingdom }

\date{\today}

\begin{abstract}
Electron pumps generate a macroscopic electric current by controlled manipulation of single electrons. Despite intensive research towards a quantum current standard over the last 25 years, making a fast and accurate quantised electron pump has proved extremely difficult. Here we demonstrate that the accuracy of a semiconductor quantum dot pump can be dramatically improved by using specially designed gate drive waveforms. Our pump can generate a current of up to 150 pA, corresponding to almost a billion electrons per second, with an experimentally demonstrated current accuracy better than 1.2 parts per million (ppm) and strong evidence, based on fitting data to a model, that the true accuracy is approaching 0.01 ppm. This type of pump is a promising candidate for further development as a realisation of the SI base unit ampere, following a re-definition of the ampere in terms of a fixed value of the elementary charge. 
\end{abstract}


\maketitle 

The AC Josephson effect and quantum Hall effect have revolutionised electrical metrology by providing stable reference standards linked to fundamental physics rather than particular artefacts \cite{zimmerman1998primer}. The resulting system of electrical units is coupled to the SI only through very difficult and time-consuming electro-mechanical experiments \cite{kibble1991present}, and recently a debate has ignited about re-defining the SI ampere in terms of the fundamental elementary charge $e$ \cite{milton2010quantum}. A re-defined ampere could be realised directly using a quantised electron pump, which generates a current by moving single electrons rapidly and in a controlled way. The electron pump accepts a periodic input signal at a repetition frequency $f$, and transports an integer number $n$ of electrons between source and drain leads for each cycle of the input to yield a current $I_{\rm P}=nef$. Application as a quantum standard of electric current requires a pump having a combination of accuracy, simplicity of operation and ability to generate a reasonably high current $I_{\rm P}\geq 100$pA \cite{zimmerman2003electrical}. In the past two decades, many pump technologies have been investigated in pursuit of this goal. These include chains of sub-micron normal-metal tunnel junctions \cite{keller1996accuracy,keller1999capacitance}, quantum dots driven by a surface acoustic wave \cite{shilton1996high}, and normal metal / superconductor turnstiles \cite{pekola2007hybrid,maisi2009parallel}. Although $0.015$~ppm accuracy has been demonstrated for $I_{\rm P}\approx1$~pA \cite{keller1996accuracy}, and  $I_{\rm P}\approx500$~pA with around $100$~ppm accuracy \cite{janssen2000accuracy}, none of these technologies have yet demonstrated the required combination of accuracy and large enough current output.

Our tunable-barrier electron pump [Scanning Electron Microscope (SEM) image in Fig.~1a, with associated biasing voltages also shown] consists of a conducting channel etched in a 2-dimensional electron gas (2-DEG), crossed by two metallic gates, which we denote the entrance and exit gates. Potential barriers are created in the 2-DEG below the gates by applying negative voltages $V_{\rm G1}$ and $V_{\rm G2}$ to these gates. An isolated quantum dot, holding a small number of electrons, forms between the gates. A drive signal $V_{\rm RF}(t)$ superimposed on $V_{\rm G1}$ periodically lowers the entrance barrier below the Fermi level of the source lead, picking up electrons and lifting them over the exit barrier into the drain lead. The exit barrier is kept high ($>10$~meV above the Fermi level) to suppress unwanted thermally-activated and co-tunnel transport \cite{zimmerman2004error}. A series of schematic potential diagrams in Fig.~1b illustrate the pump cycle for the case of $n=1$. The pump operates at zero source-drain bias voltage; the direction of the pump current is determined by which gate is driven with the AC signal. If the amplitude of $V_{\rm RF}$ is large enough, the probability of the trapped electron being ejected into the drain lead [frame 4 of Fig.~1(b)] is unity, and the current quantisation mainly depends on the initial part of the pump cycle (frame 2). Here, the relatively large number of electrons initially trapped by the rising entrance barrier is reduced to just one by a cascade of `back-tunnel' events whereby excess electrons return to the source lead (red arrow) \cite{fujiwara2008nanoampere,kashcheyevs2010universal}. Due to the Coulomb charging energy of the dot, the back-tunnel probability for the last electron is many orders of magnitude less than for the second electron, so the convergence on the 1-electron state is robust and insensitive to details of the potential barrier shape. The GaAs pump has an important advantage compared to other types of electron pump, which is that the quantisation accuracy can be improved with an external tuning parameter, the perpendicular magnetic field $B$ \cite{wright2008enhanced,kaestner2009single,leicht2011generation,wright2011single,fletcher2011stabilisation}. The mechanism for this improvement is not fully understood, but the additional confinement of electrons within the dot due to the magnetic field plays an important role \cite{fletcher2011stabilisation}. 

\begin{figure}
\includegraphics[width=8cm]{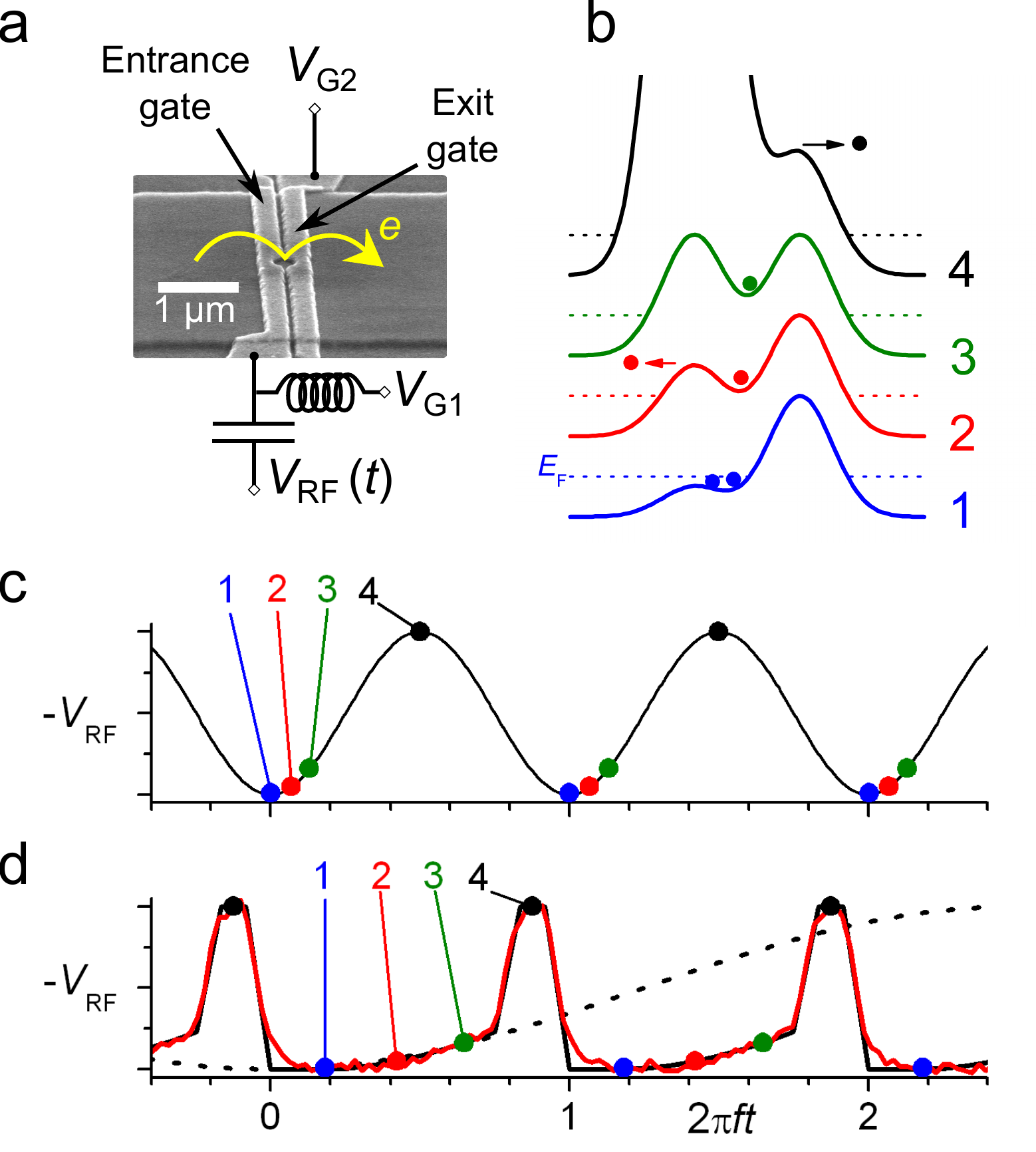}
\caption{\textsf{\textbf{Sample, pump mechanism and gate drive waveforms. (a)} SEM image of a pump similar to the one used in this study. The etched 2-DEG channel runs from left to right, and the two metallic gates (light grey in the image) cross the channel. \textbf{(b)} Schematic diagrams of the potential along the channel during four phases of the pump cycle: (1): loading, (2): back-tunneling, (3): trapping, (4): ejection. One cycle transports an electron from the left (source) to the right (drain) lead. \textbf{(c,d)} The two types of gate drive waveforms $V_{\rm RF}(t)$ used in this study: sine waves (c) and shaped pulses (d). Numbered points approximately indicate the corresponding stages in the pump cycle shown in panel (b). In panel (d), the black line shows the pulse shape programmed into the AWG, and the red line shows the actual waveform measured with a fast sampling oscilloscope after passing through a co-axial line similar to the one in the cryogenic measurement probe. The dotted line shows a sine wave of frequency $f/5$.}}
\label{fig:fig1}
\end{figure}

Previous high-resolution measurements \cite{blumenthal2007gigahertz,giblin2010accurate} used sine wave drive for $V_{\rm RF}(t)$. The sine wave cycle is shown in Fig.~1c with numbered points approximately indicating the stages of the pump cycle shown in Fig. 1b. These measurements could not be extended above $f\approx350$~MHz because the quantised plateau degraded with further increase in $f$ \cite{giblin2010accurate}. The general mechanism for this is not currently understood, although clear signatures of back-tunneling due to non-adiabatic excitation at high $f$ have been seen in some samples \cite{kataoka2011tunable}. To extend the pump operation range to higher frequencies, we developed a technique using an arbitrary waveform generator (AWG) to generate a custom $V_{G1}(t)$ waveform which takes into account the time-dependent electron tunelling dynamics. To design the waveform, firstly we note, as shown schematically in Fig.~1c, that, for sine wave drive, most of the cycle time is used to raise the trapped electron over the exit barrier, and following ejection, lower the empty dot in readiness for the next cycle. This is a consequence of the height of the exit barrier and the large amplitude ($\approx 0.5$~V) of the entrance barrier modulation. Secondly, we postulate that the increased error rate at high frequency is related to the rate at which the entrance barrier is raised during the early part of the pump cycle, when back-tunneling occurs. Our AWG waveform, shown in Fig.~1d, was designed with a small initial $dV_{\rm G1}/dt$, speeding up for subsequent parts of the pump cycle. Our technique has some similarity to earlier work employing trapezoidal gate pulses with variable rise-time \cite{fujiwara2008nanoampere} and we will return to this point later in the discussion section. The output of the AWG measured through a similar transmission line to the one used in the experiment showed a slight smearing of the actual $V_{G1}(t)$ applied to the entrance gate (red line in Fig.~1d), but this does not affect the important part of the cycle, which contains only low frequency components. The `slow' part of the waveform with repetition frequency $f$ is approximately a segment of sine wave of frequency $f/5$ (dotted line), enabling us to slow down the electron capture process by a factor of 5. Thus we expect the pump driven with the AWG pulse waveform at frequency $f$ to have the same characteristics as it would have with sine wave drive at frequency $f/5$.

Using the AWG waveform, and employing a more accurate current measurement system (described in detail in the methods section), we demonstrate a semiconductor pump generating currents up to $150$~pA accurate to better than $1.2$~ppm, the limit set by our measurement uncertainty. Theoretical analysis predicts that the true pump accuracy is between one and two orders of magnitude better than our experimental uncertainty, making this type of pump an extremely promising candidate for development as a future quantum representation of the ampere.

\subsection*{RESULTS}
\subsubsection{Accuracy of the pump.} 
We cooled our pumps to $300$~mK and measured the pump current $I_{\rm P}$ using a technique described in more detail in the methods section. All high-resolution data in Figs.~2,3,4 and 6 were obtained in a perpendicular magnetic field $B=14$~T. Most measurements reported in this paper were obtained on a single sample, denoted sample 1. Some measurements were made on an additional sample, denoted sample 2. Data is for sample 1 unless stated otherwise. Figure 2 shows a measurement of the pump current on the $n=1$ plateau as the exit gate voltage is varied. The 9 red data points were obtained in 3 experimental runs spread over 4 different days, and their mean and standard deviation are shown by the horizontal pink line (with error bar indicating the 1.2 ppm systematic uncertainty), and grey box respectively. This data demonstrates the stability, plateau flatness and absolute accuracy of the pump at the 1~ppm level. The accuracy can be seen by comparing the mean, with its associated uncertainty, with the value of $ef$ indicated by the blue dotted line. We used the 2006 CODATA value of $e$, which has an uncertainty, insignificant for this measurement, of $0.025$~ppm \cite{mohr2008codata}. On the plateau, the difference $I_{\rm P}-ef$ is $(-0.077\pm0.18)$ fA, or $(-0.51\pm1.2)$ ppm. 

\begin{figure}
\includegraphics[width=9cm]{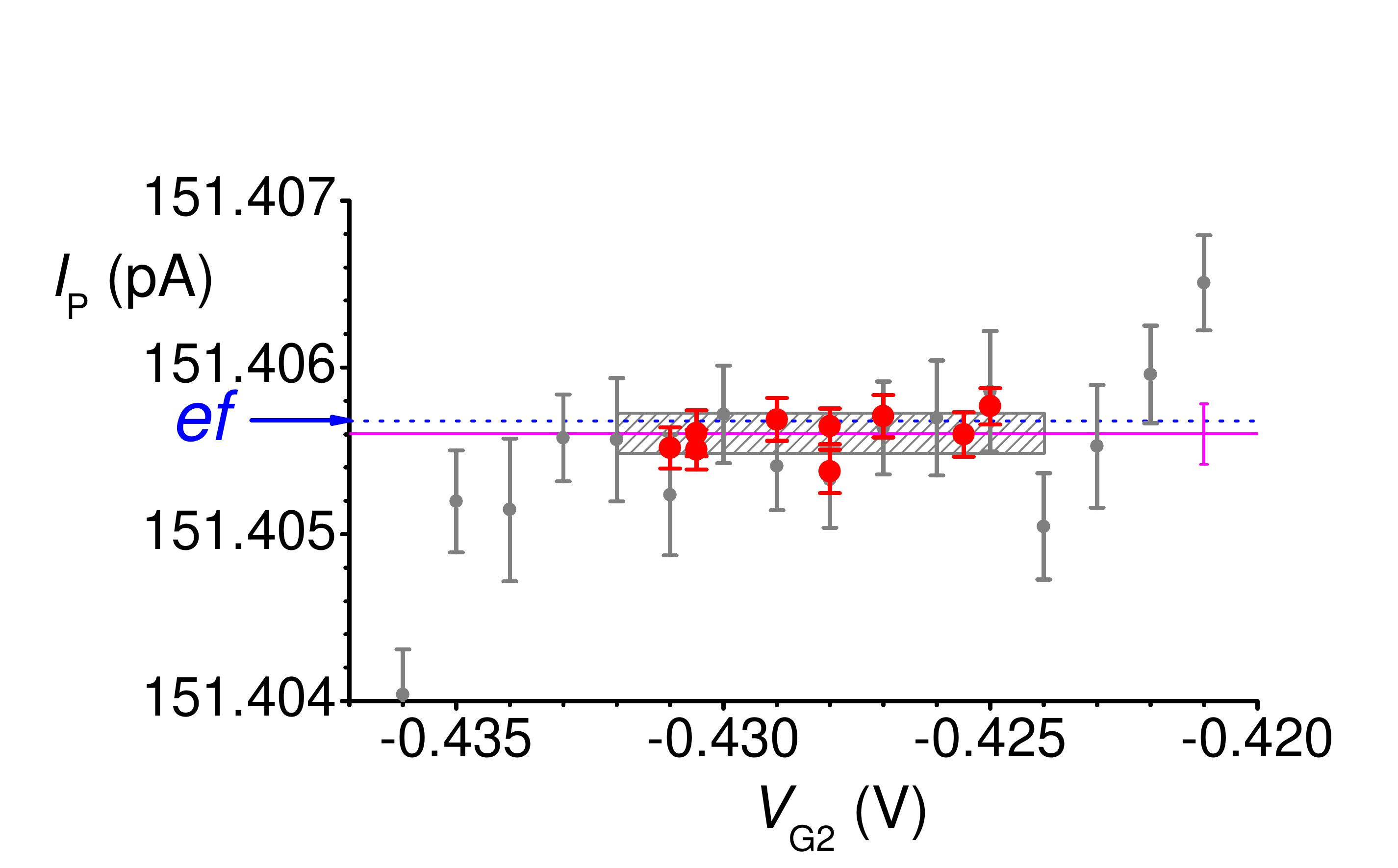}
\caption{\textsf{\textbf{High-accuracy data.} High-resolution measurement of the pump current as a function of exit gate voltage on the $n=1$ plateau with $f=945$~MHz, $B=14$~T obtained by averaging over $17$ (small grey points) and $96$ (large red points) on-off cycles (Fig. 7 and methods section). Error bars show the random uncertainty $U_{\text{R}}$. The horizontal dotted line shows the expected current corresponding to exactly one electron pumped for each pump cycle, and the horizonal solid line shows the mean of the $9$ red points, with error bar indicating the systematic uncertainty $U_{\text{S}}$.}}
\label{fig:fig2}
\end{figure}

\subsubsection{Quantisation improvement using AWG drive.} 
The dramatic effect of using AWG pulse drive on sample 1 is shown in more detail in Fig.~3. First, we made high resolution measurements of the pump current as a function of $V_{\rm G2}$, using conventional sine wave drive, shown in Fig.~3a. The quantity plotted is the fractional deviation of pump current from $ef$ in parts per million, defined as $\Delta I_{\rm P}\equiv 10^6(I_{\rm P}-ef)/ef$. At $400$~MHz a broad plateau is seen, accurately quantised within the uncertainty of $\sim3$~ppm, which is dominated by the random uncertainty $U_{\text{R}}$ for this combination of $f$ and averaging time. When the frequency is increased to $630$~MHz, the plateau becomes narrower, although it remains quantised over a finite range of $V_{\rm G2}$. Further increase in $f$ results in a catastrophic loss of quantisation. Figures 3b and 3c illustrate how the quantisation can be recovered by switching to the AWG pulse waveform illustrated in Fig. 1d. At $630$~MHz the pulse waveform results in a $\approx\times2$ increase in the width of the plateau compared to the sine wave drive, but at $945$~MHz, where no plateau was present on this expanded current scale with sine wave drive, the pulse waveform completely restores accurate quantised pumping. Data for sample 2, at $f=630$~MHz, is shown in Fig. 4a. This sample did not exhibit as good a quantised plateau as sample 1, illustrated by the lack of a plateau using sine wave drive. However, the plateau could be progressively recovered using AWG waveforms with slower rising edges, illustrated in Fig. 4b. Waveform AWG1 (AWG2) had a rising edge corresponding to a sine wave of frequency $315$ ($218$)~MHz.

\begin{figure}
\includegraphics[width=9cm]{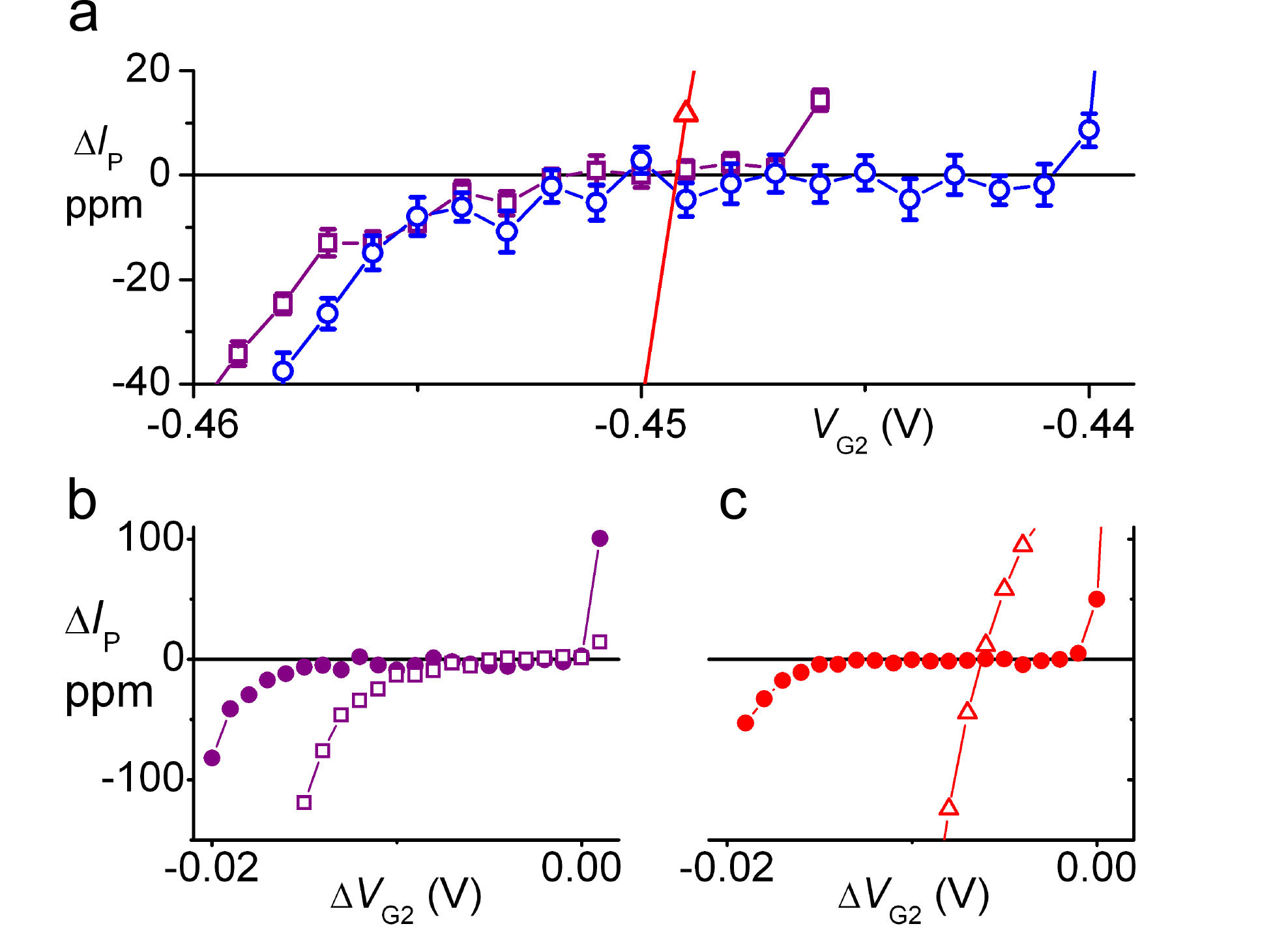}
\caption{\textsf{\textbf{Effect of AWG pulse waveform on pump performance for sample 1. (a)} Fractional deviation of pump current from $ef$ in parts per million, as a function of exit gate voltage, using sine wave drive at $400$~MHz (circles), $630$~MHz (squares) and $945$~MHz (triangles). \textbf{(b) and (c)} Pump current as a function of offset exit gate voltage $\Delta V_{\rm G2}=V_{\rm G2}-V_{\rm G2,H}$, where $V_{\rm G2,H}$ is the high-voltage edge of the $n=1$ plateau, at $630$~MHz (panel b) and $945$~MHz (panel c). Open (solid) points indicate sine wave (AWG pulse) drive.}}
\label{fig:fig3}
\end{figure}

\begin{figure}
\includegraphics[width=9cm]{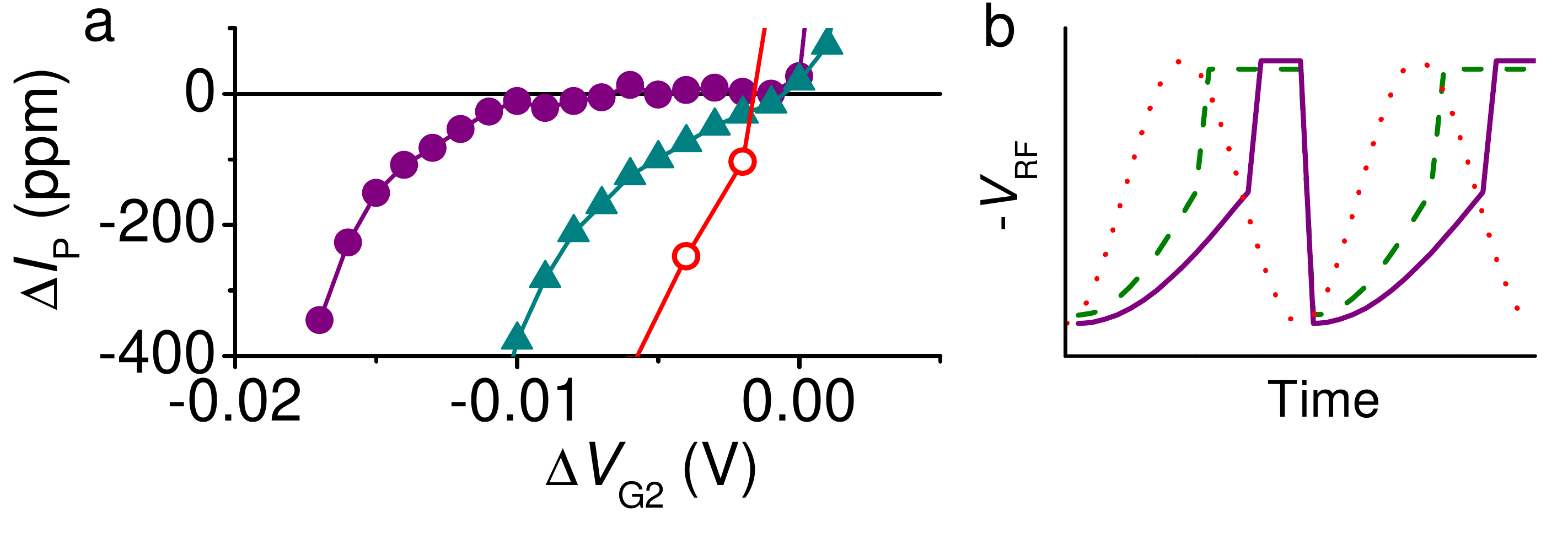}
\caption{\textsf{\textbf{Effect of AWG pulse waveform on pump performance for sample 2. (a)} Fractional deviation of pump current from $ef$ in parts per million, as a function of normalised exit gate voltage at $630$~MHz, using sine wave drive (open circles), waveform AWG1 (closed triangles) and waveform AWG2 (closed circles). \textbf{(b)} The three drive waveforms used to obtain the data in panel (a). Dotted line: sine wave, dashed line: AWG1, solid line: AWG 2. The overall amplitude of the waveforms is $\sim0.8$~Vpp.}}
\label{fig:fig4}
\end{figure}

\subsubsection{Fitting to the decay cascade model.} The improvement using the AWG pulse drive can be demonstrated more quantitatively by considering theoretical predictions for the overall shape of the $I_{\rm P}(V_{\rm G2})$ plateau. In Fig.~5a we plot the measured average number of electrons pumped per cycle $\langle n\rangle\equiv I_{\rm P}/ef$, for four combinations of magnetic field and pumping frequency, using sine wave drive. In contrast to the high-resolution data of Figs.~2b, 3 and 4, these data were taken over a wide range of $V_{\rm G2}$, with relatively low current resolution, to show the full transition $n=0\rightarrow1$. As $B$ is increased, the $n=1$ plateau becomes wider and the transitions between plateaux become sharper, resulting in improved current accuracy on the plateau, while increasing $f$ has the opposite effect. The $\langle n\rangle(V_{\rm G2})$ data are fitted to an analytical formula derived from the decay cascade model \cite{fujiwara2008nanoampere,kashcheyevs2010universal}: \begin{equation} \langle n\rangle_{\rm FIT}=\displaystyle\sum\limits_{m=1}^2 \exp(-\exp(-aV_{\rm G2}+\Delta_{m}))
\end{equation} where $a, \Delta_{1}, \Delta_{2}$ are fitting parameters. The exponential approach to the $n=1$ quantised plateau, as $V_{\rm G2}$ is made more positive, is understood as an exponentially decreasing probability of the electron tunneling back out of the dot into the source lead, while excess electrons tunnel back out with essentially unit probability (Fig.~1b). The parameter $\delta_{2}=\Delta_{2}-\Delta_{1}$ can be used as a practical figure of merit for evaluating the accuracy of a given pump, because there is a simple relationship between $\delta_{2}$ and the predicted pump error $\varepsilon_{\rm P}\equiv1-\langle n\rangle_{\rm FIT}$ at the point of inflection in $I_{\rm P}(V_{\rm G2})$ \cite{kashcheyevs2010universal}. The point of inflection can be found from a relatively fast, low-resolution measurement. The fit line and its derivative for the 1~GHz data are shown in Fig.~5b with axes expanded to show $\varepsilon_{\rm P}$. 

\begin{figure}
\includegraphics[width=9cm]{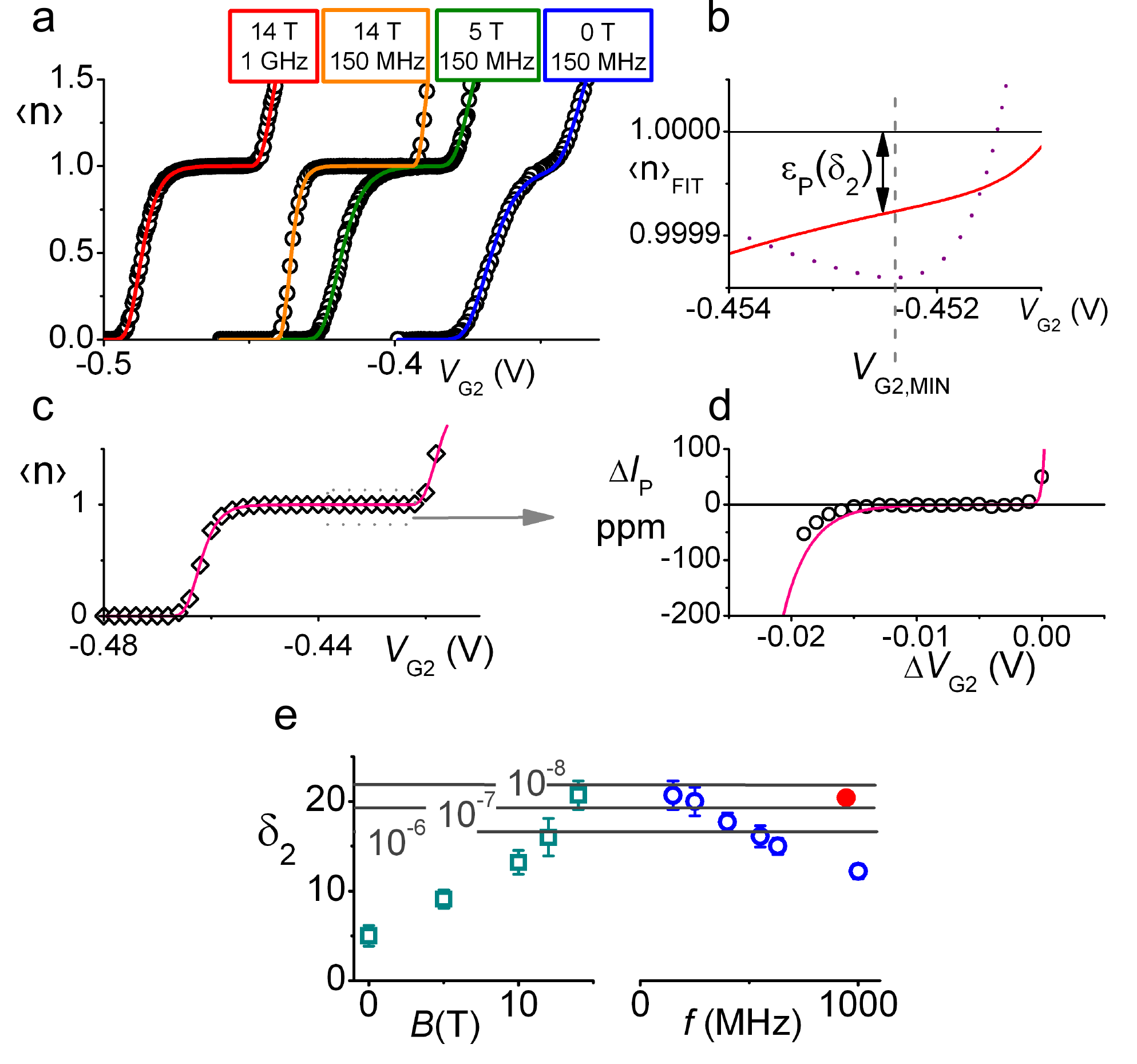}
\caption{\textsf{\textbf{Comparison of data with the decay cascade model. (a)} Points: average number of pumped electrons $\langle n\rangle\equiv I_{\rm P}/ef$ as a function of exit gate voltage for four combinations of magnetic field and pump frequency, using sine wave drive. Solid lines: $\langle n\rangle_{\rm FIT}$ obtained by fitting the data to equation (1). Values of the fitting parameter $\delta_2$ are (left-right) $12.2, 20.7, 9.1$ and $4.6$. \textbf{(b)} Solid line: close-up of the $\langle n\rangle_{\rm FIT}(V_{\rm G2})$ line for the 1~GHz data in (a). Dotted line: derivative $d\langle n\rangle_{\rm FIT}/dV_{\rm G2}$ (arb. units) which is minimised at gate voltage $V_{\rm G2,MIN}$. The theoretically predicted pump error $\varepsilon_{\rm P}\equiv1-\langle n\rangle_{\rm FIT}$ is defined for $V_{\rm G2}=V_{\rm G2,MIN}$. \textbf{(c)} Points: average number of pumped electrons as a function of exit gate voltage obtained from a quick low-resolution measurement ($0.4$~s / data point) for AWG drive at $945$~MHz. Line: $\langle n\rangle_{\rm FIT}$ obtained by fitting the data to equation (1). \textbf{(d)}: comparison between the fit from (c) (line) and the high-resolution data using AWG drive at $945$~MHz from Fig.~3(c) (points) plotted on an offset gate voltage scale. \textbf{(e)} Fitting parameter $\delta_2$ extracted from fits similar to those shown in (a). Horizontal dotted lines show the values of $\delta_2$ corresponding to predicted pumping errors $\varepsilon_{\rm P}$ of $1, 0.1$ and $0.01$~ppm. The left panel shows $\delta_2$ as a function of $B$ for $f=150$~MHz, sine wave drive, and the right panel shows $\delta_2$ as a function of $f$ for $B=14$~T. Open circles: sine wave drive, closed circle: AWG drive.}}
\label{fig:fig5}
\end{figure}

\subsubsection{Comparison of data with decay cascade model.} 
Our high resolution measurements enable us to investigate the agreement between $ef\langle n\rangle_{\rm FIT}$ obtained from low-resolution data, and the actual value of the pump current on the plateau. Figs.~5c and 5d show a fit to low-resolution data, and the comparison with the high-resolution data, using AWG drive at $945$~MHz. It can be seen that the fit slightly under-estimates the width of the plateau, and consequently $\varepsilon_{\text{P}}$ is an over-estimate of the error at $V_{\text{G2}}=V_{\text{G2,MIN}}$, consistent with earlier, lower-resolution, measurements \cite{giblin2010accurate}. The other high-resolution data sets contributing to our assessment of the pump accuracy posessed similar properties; the fit always under-estimated the true plateau flatness. Consequently, we consider $\varepsilon_{\text{P}}$ as a lower-limit (in other words, a pessimistic estimate) to the predicted pump accuracy, which for the $945$~MHz data is approaching 0.01 ppm. 

In Fig.~5e we show $\delta_{2}$ as a function of $B$ deduced from the experimental data measured with sine wave drive at $f=150$~MHz (left panel) and as a function of $f$ measured at $B=14$ T with sine drive (right panel, open circles) and AWG pulse drive (right panel, filled circle). Horizontal lines show the thresholds for obtaining predicted accuracies of 1, 0.1 and 0.01~ppm. For $f\approx 1$ GHz using pulse AWG drive and $B=14$~T, we predict an accuracy better than 0.1 ppm. The value of $\delta_2$ extracted from the AWG pulse data at $f=945$~MHz roughly matches that for $f\sim200$~MHz using sine drive. This is in good agreement with the $\times5$ scaling of the time scale expected from the design of the pulse waveform. The data of Fig.~5e together with Figs 3 and 4 support our assumption that the pump errors at high frequency originate during the first part of the pump cycle, and can be minimised by slowing down this part of the cycle at the expense of other parts.

\begin{figure}
\includegraphics[width=7cm]{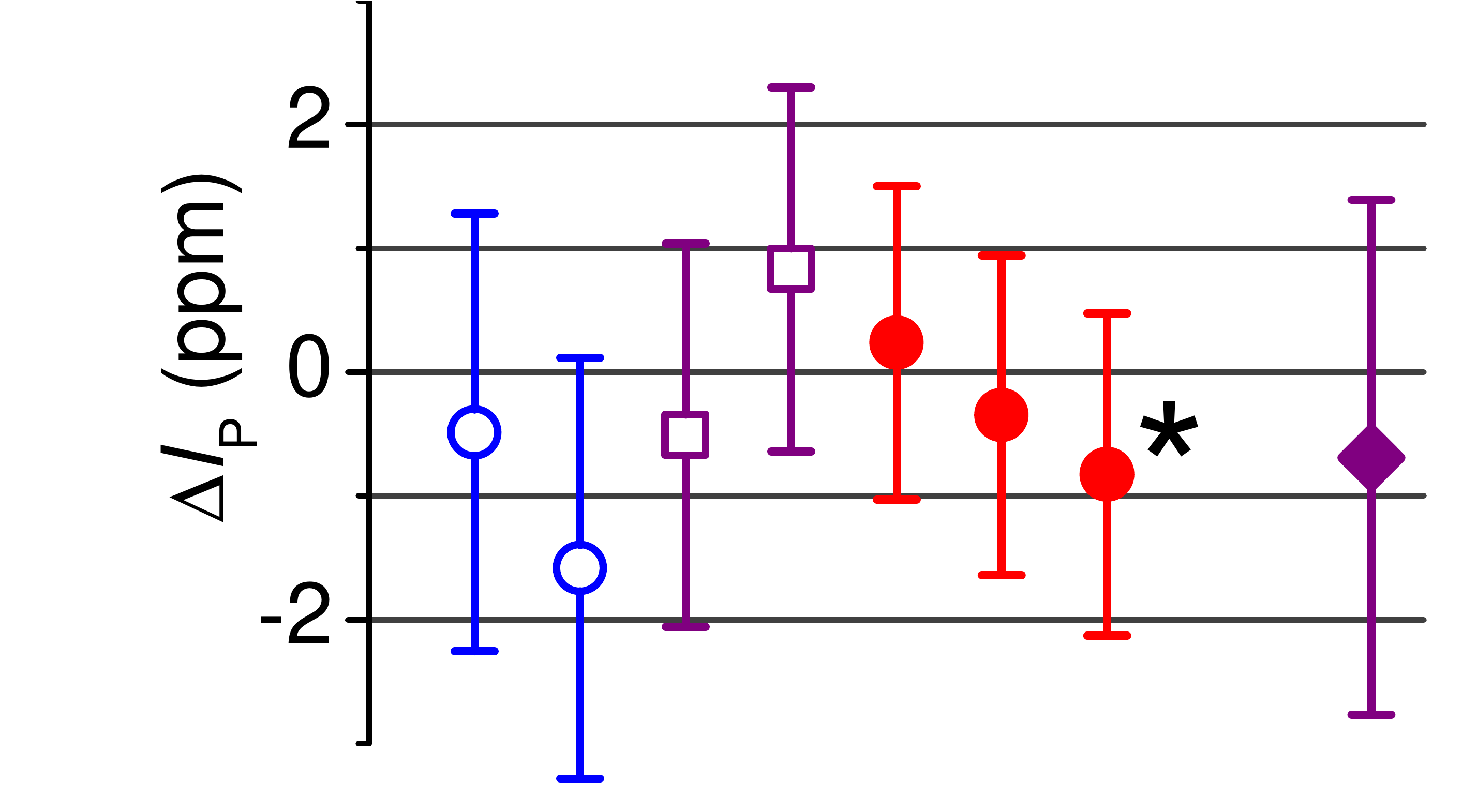}
\caption{\textsf{\textbf{Summary of experimentally demonstrated pump accuracy} Fractional deviation of pump current from $ef$ in parts per million, averaged from seven experimental runs on sample 1 at $400$~MHz, sine (open circles), $630$~MHz, sine (open squares) and $945$ MHz, AWG (closed circles), and one experimental run on sample 2 at $630$~MHz, AWG (closed diamond). Error bars show the total uncertainty $U_{\rm T}$. The data point marked with a * was measured with $V_{\rm G1}$ shifted by $10$ mV away from the plateau centre.}}
\label{fig:fig6}
\end{figure}

\subsubsection{Summary of pump accuracy.} 
In Fig.~6 we summarise the main experimental finding of this study. This figure shows the pump current averaged over the $V_{\rm G2}$ plateau for 7 experimental runs at three frequencies for sample 1, and a single run for sample 2. The total measurement time is around 10 hours per data point. We see that any deviation of the pump current from $ef$ is less than our total measurement uncertainty $U_{\rm T}=\sqrt{U_{\rm S}^2 + U_{\rm R}^2}$, which is as low as 1.2~ppm for the highest pump frequencies where the larger signal-to-noise ratio reduces the contribution of $U_{\rm R}$ to a negligible level. A full investigation of the invariance of the pump current as a function of the additional control parameters $V_{\rm G1}$, the source-drain bias voltage and the amplitude of $V_{\rm RF}$ is beyond the scope of this study, but an initial investigation into the invariance of $I_{\text{P}}(V_{\rm G1})$ was undertaken. The data point in Fig. 6 marked with a star was obtained with $V_{\rm G1}$ shifted by $10$ mV away from the plateau centre. No significant deviation from $ef$ was seen following this adjustment. 

\subsection*{DISCUSSION}

We have demonstrated that the pumping errors at high frequency can be reduced by many orders of magnitude by the use of specially-designed gate drive waveforms with a slow initial rising edge. The data of Figs. 3 and 4 support our postulate that errors at high pumping frequency arise during the initial phase of the pump cycle, when the number of electrons in the dot is reduced by the back-tunnel cascade. However, the reason for this is unclear. Indeed, the back-tunnel model \cite{fujiwara2008nanoampere,kashcheyevs2010universal} does not predict any characteristic time-scale for the decay cascade as a consequence of assuming a strictly exponential dependence for the tunnel rates on $V_{\rm{G1}}$. Instead, it predicts a shift of the plateau to more negative $V_{\rm{G2}}$ as the rise time of $V_{\rm{G1}}$ is decreased. As already noted, one experiment on silicon MOSFET pumps provides support for the simple back-tunnel model: as the rise time of trapezoidal gate drive pulses was decreased from $20$~ns to $2$~ns, the plateaus shifted to more negative values of the top gate which controlled the dot depth, but the plateau shape did not change \cite{fujiwara2008nanoampere}. We do not believe this result contradicts our own findings. We have observed the effect of decreasing rise time through the degredation of initially very flat plateaus measured to high resolution. In contrast, the plateaus presented in the variable rise-time experiment of ref. \cite{fujiwara2008nanoampere} were clearly not flat even at the longest rise-time investigated. It is likely that the plateau flatness in this case was not limited by the rise-time, and if much shorter rise times than $2$~ns had been employed, a further reduction in the plateau flatness would have been seen, in agreement with our experiments. Understanding the upper frequency limit of accurate pumping in these devices requires more accurate physical models of the pumping mechanism, for example by incorporating a physically realistic model of the time-dependence of tunnel rates into the back-tunnel model, or including the effect of non-adiabatic excitations \cite{kataoka2011tunable}.

This is the first time an electron pump current has been \textit{directly} compared to a reference current generated outside the pump cryostat with an uncertainty of order 1~ppm. In experiments using metallic pumps, a small pump current $I_{\rm{P}}\lesssim0.5$pA was compared to units of voltage, capacitance, and time by charging a capacitor located in the same cryostat as the pump, and measuring the resulting voltage change across the capacitor \cite{keller1999capacitance,camarota2012electron}. The lowest uncertainty in these experiments was 0.92~ppm \cite{keller2007uncertainty}, similar to that in the present work, but with a current 300 times smaller. At the $100$ pA current level of the semiconductor pump, a direct realisation of the quantum metrological triangle \cite{piquemal2000argument} at the $10^{-7}$ uncertainty level is feasible within an averaging time of several hours. Direct calibrations of a pico-ammeter, or a micro-ammeter using a suitable high-gain current comparator \cite{rietveld2003CCC,feltin2003progress,steck2008characterization} are also possible. This makes our result a significant step towards the application of electron pumps in primary electrical metrology. We have shown that the pump current is invariant in one control paramter ($V_{\text{G2}}$) and equal to $ef$ to within $1.2$~ppm. Future experiments will investigate in detail the invariance of the current across the additional control parameters, and over a range of samples. 

Another class of experiment uses mesoscopic charge detectors capactively coupled to part of the pump circuit to count transport errors at the single electron level \cite{keller1996accuracy}. These experiments enable measurements of much smaller error rates than are possible by averaged current measurements such as those reported in this paper. Furthermore, they provide a direct measure of the pump error which is competely independent of the performance of any reference standard. If the electron pump is to be used as the current leg of a metrological triangle experiment, charge-counting accuracy tests are important because they allow the experiment to distinguish between errors in the number of transported charge quanta, and a more fundamental correction to the value of the charge on each quantum \cite{keller2008current}. Error-counting experiments at the GHz frequencies employed in semiconductor pumps are technically challenging due to the limited bandwidth of detectors, but some initial progress has already been made \cite{Fricke2011quantized,yamahata2011accuracy}, and these experiments will continue to be persued. 

In summary, we have demonstrated a milestone in single-electron control, by transferring $\approx 10^9$ electrons per second through a semiconductor quantum dot with a resulting current accuracy better than 1.2~ppm. By using shaped pulses to drive the control gate, we are able to limit errors and operate the pump accurately at higher frequency than was previously possible. These results are extremely encouraging for the development of a quantum current standard, re-definition of the unit ampere, and more generally for precise high speed control of electrons in semiconductor devices.

\subsection*{METHODS}

\subsubsection{Sample and Cryostat.} 
We fabricated our pumps on GaAs/Al$_{x}$Ga$_{1-x}$As wafers using standard techniques: wet-etching to define the channel, and electron-beam lithography for the metallic gates \cite{blumenthal2007gigahertz}. The samples were cooled in a sorbtion-pumped helium-3 refrigerator with a base temperature of $300$~mK. The special gate pulses were generated using a $12$~GS/s Tektronix 7122B arbitrary waveform generator.

\subsubsection{Current measurement system.} 
Conventional electronic pico-ammeters are limited in practice to $\approx100$~ppm accuracy, even following a calibration, due to drift in the gain of the pre-amp stage. To achieve part-per million accuracy, we generated a reference current $I_{\rm R}$ with opposite polarity to $I_{\rm P}$ by applying a voltage $V_{\rm C}$ across a temperature controlled ($26.57 \pm0.02~\tccentigrade$) $1$~G$\Omega$ standard resistor $R$ (circuit in Fig.~7a). A room-temperature current pre-amplifier, with transimpedance gain $\approx10^{10}$~V/A calibrated to 0.1~\% accuracy, measured the small difference between the pump and reference currents. To remove offset currents and thermoelectric potentials, $V_{\rm C}$ and $V_{\rm RF}$ were synchronously switched on and off with a cycle time of 150~s. The differences in current and voltage, $\Delta I$ and $\Delta V$ respectively, were recorded for each cycle as indicated in the section of raw data shown in Fig.~7b. The pump current is given by $I_{\rm P}=\Delta V/R + \Delta I+(r/R)\Delta I$, where $r\approx 10$~k$\Omega$ is the input resistance of the pre-amp. Because $|\Delta I|/I_{\rm P} ,  r/R \lesssim10^{-4}$, the final term can be neglected and the pump current is determined by the voltmeter reading, the value of $R$, and the small residual current $\Delta I$ with the pre-amp gain calibration contributing a small ($<0.1$~ppm) correction. The voltmeter and resistor were calibrated, via some intermediate steps, against the AC Josephson effect and quantum Hall resistance respectively. Formally we have measured current in 1990 units, using the agreed values of the von Klitzing and Josephson constants $R_{\text{K-90}}$ and $K_{\text{J-90}}$. The uncertainties of $R_{\text{K}}$ and $K_{\text{J}}$ in SI units are sufficiently small, less than $0.1$~ppm, that we can quote our result in SI units witout a significant increase in uncertainty. The resistor was calibrated at a voltage of $10$~V, with additional calibrations performed at higher voltages to check for possible power-coefficient effects as shown in Supplementary Figure S1. No power co-efficient was found in the voltage range $10$~V $\leq V\leq 100$~V. Below $10$~V, the resistor dissipates less than $100$~nW and power co-effiecient effects are expected to be insignificant. 

\begin{figure}
\includegraphics[width=9cm]{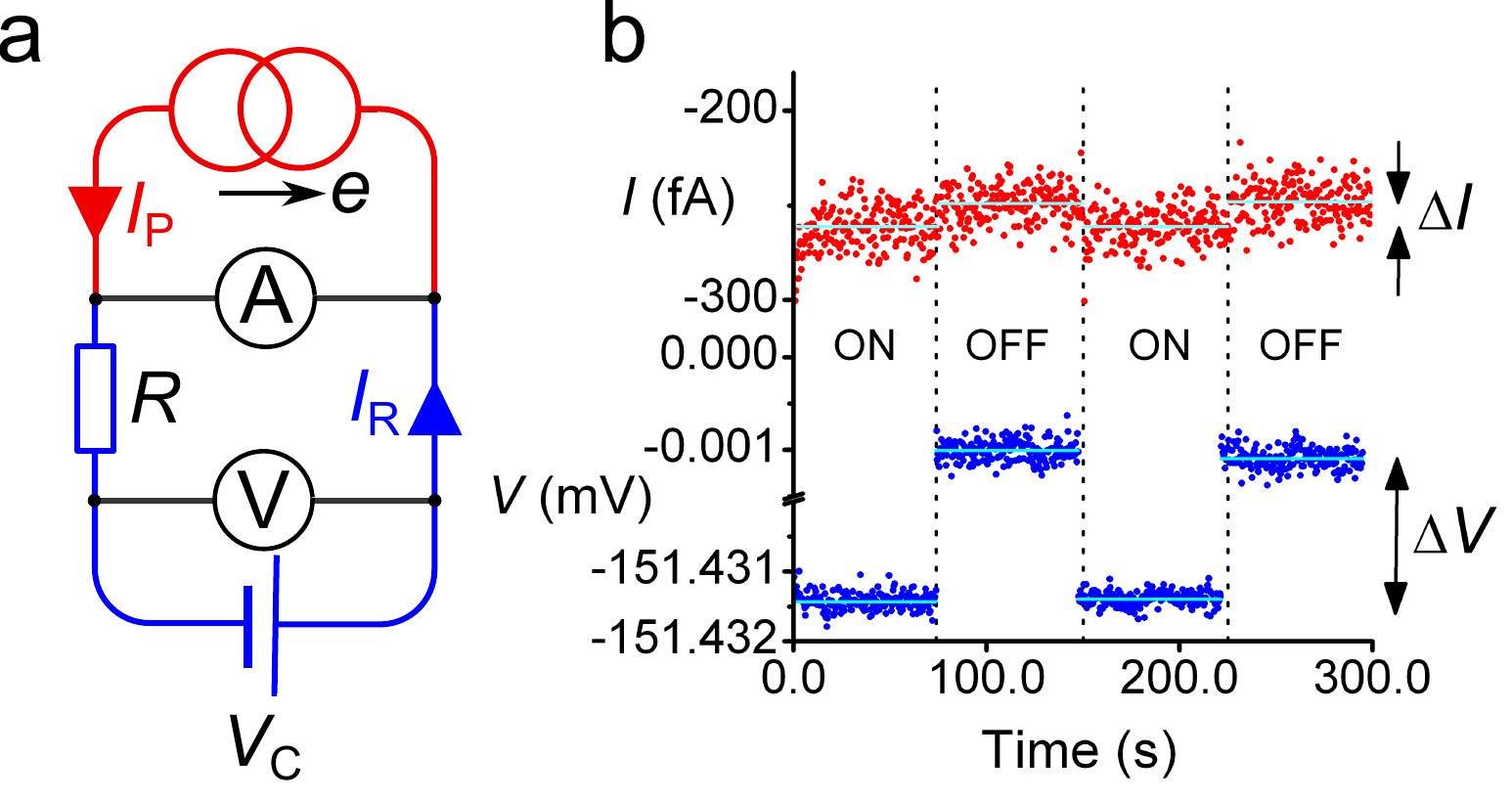}
\caption{\textsf{\textbf{Measurement circuit and raw data. (a)} Circuit used for accurate measurement of the pump current. The pump is depicted as a current source. \textbf{(b)} A short section of raw ammeter (upper panel, red points) and voltmeter (lower panel, blue points) data from a measurement run pumping with $f=945$~MHz, showing two on-off cycles. Horizonal lines show the means of each data segment, ignoring the first 20 readings to reject transient effects. The offset current $\approx -250$~fA results from the small stray bias voltage present at the pre-amp input, driving a current in $R$.}}
\label{fig:fig7}
\end{figure}

\subsubsection{Uncertainty.} 
The total systematic uncertainty $U_{\text{S}}$ in our measurement of the pump current is 1.2~ppm, dominated by the 0.8~ppm systematic uncertainty in the calibration of the resistor. The $2\sigma$ calibration uncertainty quoted for 1~G$\Omega$ resistors at NPL is $1.6$~ppm. The relevent entry on the peer-reviewed calibration and measurement capability (CMC) data base is at \cmcdata. All unceratinties quoted in this paper are $1\sigma$ and have been rounded to the nearest $0.1$~ppm. A full break-down of the uncertainty is shown in Supplementary Table S1. Calibrations of $R$ were performed at regular intervals over several years to characterise the small drift, which is common in artefact standards: $R=1~000~051~260\pm800~\Omega$ on 27/4/2011, with a drift correction of $11~\Omega/$day. The maximum drift correction applied for pump measurements before and after the calibration date was $\approx0.5$~ppm. The random uncertainty $U_{\rm R}$ is dominated by the Johnson current noise in $R \approx 4$~fA$/\sqrt{\rm Hz}$. Thus, 1~ppm resolution of $I_{\rm P}\approx 100$~pA requires averaging times of the order of 1~hour.

This research was supported by the UK Department for Business, Innovation and Skills, the European Metrology Research Programme (grant no. 217257) and the UK EPSRC.

\vspace{10pt}
Author contributions: S.P.G. Designed and calibrated the measurement system, M.K. Designed the pulse drive technique, J.D.F. designed the measurement probe, S.P.G. and M.K. performed experiments and analysed data, M.K designed the sample, P.S. performed and supervised sample fabrication, D.A.R. and I.F. provided GaAs wafers, J.P.G. and G.A.C.J. performed electron beam lithography, T.J.B.M.J. contributed project leadership and supervision, S.P.G. wrote the paper with extensive comments / revisions contributed by M.K. and J.D.F.

Author information: The authors declare no competing financial interests. Correspondence and requests for materials should be addressed to S.P.G. (stephen.giblin@npl.co.uk).

\end{document}